\documentstyle[epsfig,12pt]{article}
\newcommand{\be}{\begin{eqnarray}}
\newcommand{\ee}{\end{eqnarray}}

\def\lsim{\mathrel{\rlap{\lower4pt\hbox{\hskip1pt$\sim$}}
    \raise1pt\hbox{$<$}}}               
\def\gsim{\mathrel{\rlap{\lower4pt\hbox{\hskip1pt$\sim$}}
    \raise1pt\hbox{$>$}}}               

\def\dslash{\protect{\slash \hskip-7pt}}  

\begin{document}

\vspace{1cm}

\rightline{\Large{Preprint RM3-TH/01-1}}

\vspace{1cm}

\begin{center}

\LARGE{Electromagnetic form factors in the light-front formalism and the Feynman triangle diagram: spin-0 and spin-1 two-fermion systems\footnote{\bf To appear in Physical Review D.}}\\

\vspace{1cm}

\large{Dmitri Melikhov$^*$ and Silvano Simula$^{**}$}\\

\vspace{0.5cm}

\normalsize{$^*$Institute for Theoretical Physics, Philosophenweg 16, D-69120, Heidelberg, Germany\\$^{**}$Istituto Nazionale di Fisica Nucleare, Sezione Roma III, Via della Vasca Navale 84, I-00146 Roma, Italy}

\end{center}

\vspace{1cm}

\begin{abstract}

\noindent The connection between the Feynman triangle diagram and the light-front formalism for spin-0 and spin-1 two-fermion systems is analyzed. It is shown that in the limit $q^+ = 0$ the form factors for both spin-0 and spin-1 systems can be uniquely determined using only the {\em good} amplitudes, which are not affected by spurious effects related to the loss of rotational covariance present in the light-front formalism. At the same time, the unique feature of the suppression of the pair creation process is maintained. Therefore, a physically meaningful one-body approximation, in which all the constituents are on their mass-shells, can be consistently formulated in the limit $q^+ = 0$. Moreover, it is shown that the effects of the contact term arising from the instantaneous propagation of the active constituent can be canceled out from the triangle diagram by means of an appropriate choice of the off-shell behavior of the bound state vertexes; this implies that in case of {\em good} amplitudes the Feynman triangle diagram and the one-body light-front result match exactly. For illustrative purposes our covariant light-front approach is applied to the evaluation of the $\rho$-meson elastic form factors.

\end{abstract}

\vspace{1cm}


\vspace{0.25cm}


\newpage

\pagestyle{plain}

\section{Introduction}

\indent Theoretical description of bound states in terms of their constituents is one of the key issues of particle physics. The information on the structure of bound states is contained in the form factors which describe the properties of bound states as seen by various kinds of probes. The goal of the theory is to calculate these form factors as functions of the squared momentum $q^2 = q \cdot q$, transferred to the bound state, from the known dynamics of the constituent interactions. 

\indent The general approach to this problem is well-known \cite{IZ}: one should consider a three-point Green function  given by the triangle diagram with the full constituent propagators, constituent interaction vertex and the Bethe-Salpeter wave functions of the bound states, obtaining in this way the bound state form factors. This recipe is quite universal and in principle applies to any theory, but in practice the problem relies on the fact that in many interesting cases, like e. g. QCD, neither full propagators and vertexes, nor exact bound state wave functions are known. What is usually known is the first few terms of the perturbative expansion for the propagator and vertex functions, or in some cases the leading contribution to these quantities in some specific regions of values of the momentum transfer. An important example is the double-logarithmic asymptotic behavior of the quark vertex function at large momentum transfers which was obtained by a proper resummation of the leading perturbative corrections in Ref. \cite{Sudakov}. An elegant derivation of such an asymptotic behavior was carried out on the light-front in Ref. \cite{Jackiw}. 

\indent In QCD the additional difficulty comes from the fact that the formation of hadrons from quarks and gluons is a non-perturbative effect. Considerable simplifications emerge at values of $q^2$ much larger than the energy scale of the QCD confinement. In this region hadron form factors may be calculated in terms of the light-front hadron wave functions which describe the structure of hadrons in terms of quarks and gluons. An important feature of hard exclusive processes is that on the light-front the leading contribution is given by the Fock component of the hadron wave function with a minimal number of partons \cite{exclusive}. This property makes the analysis of hadron form factors at large $q^2$ treatable. A consistent treatment of the hadron wave functions in terms of quarks and gluons and a calculation of the perturbative corrections can be performed on light-front QCD (see Ref. \cite{LF-QCD}). 

\indent At intermediate and small values of the momentum transfer the situation looks much more complicated and the perturbative treatment cannot provide a consistent description of the form factors. In the partonic language this would mean that the infinite number of components of the Fock wave function in terms of quarks and gluons should be taken into account. Parton degrees of freedom turn out to be irrelevant for the description of form factors at small momentum transfers. 

\indent In this case one has to rely upon non-perturbative methods like QCD sum rules or lattice calculations. Another possibility is to use various phenomenological approaches based on the notion of constituent quarks, effective degrees of freedom which are expected to be relevant for the description of soft processes, i.e. processes at small momentum transfers. In terms of constituent quarks, hadrons are treated as relativistic few-body systems. Although so far there is no derivation of the constituent quark picture from QCD, this concept works surprisingly well in hadron spectroscopy and for the description of hadron form factors. 

\indent In this work we are interested in considering non-perturbative effects in the evaluation of bound-state form factors, adopting approximate propagators and vertexes in terms of constituent quarks, which may provide an interesting link with phenomenological potential models, like e.g. the spectroscopic quark model of the hadron structure.

\indent Since Dirac \cite{Dirac} it is known that there are three main ways to include the interaction term in the generators of the Poincar\'e group and to maximize at the same time the number of kinematical (interaction free) generators (see also Ref. \cite{Stern}). The three forms of the relativistic dynamics are: the instant form, where the interaction is present in the time component $P^0$ of the four-momentum and in the Lorentz boost operators; the point form, where all the components of the four-momentum operator depend on the interaction, and the light-front ($LF$) form, in which the interaction appears in the "minus" component of the four-momentum ($P^- \equiv P^0 - P^z$) and in the transverse rotations around the $x$ and $y$ axes, with the null-plane being defined by $x^+ \equiv t + z = 0$. An explicit construction of the interaction-dependent Poincar\'e generators was carried out in Ref. \cite{Bakamjian} in case of the instant form and subsequently generalized to the other two forms (see, e.g., Ref. \cite{LF}). Relativistic quantum models, based on the abovementioned forms of the dynamics with a fixed number of constituents, have been widely used for the investigation of hadron phenomenology, such as the electroweak properties of mesons and baryons within the framework of the constituent quark picture, or the elastic electromagnetic form factors of the deuteron viewed as a composite two-nucleon system. 

\indent However, the description of a bound state in terms of a fixed (usually minimal) number of constituents violates in general the requirement of (extended) Poincar\'e covariance. As a matter of fact, the interaction term may generally cause the creation of particle-antiparticle pairs from the vacuum (the so-called $Z$-graph), so that, e.g., the $LF$ transverse rotations may change the number of constituents (cf. Ref. \cite{Miller}). Thus, the use of relativistic wave functions with a fixed number of constituents may be consistent only with the transformations associated to the given kinematical subset of Poincar\'e generators, and moreover it is expected to be physically meaningful only in reference frames where the pair creation process can be suppressed (cf. Ref. \cite{Brodsky}).

\indent A further source of breaking of Poincar\'e covariance is the use of approximate current operators and in this work we will limit ourselves to the case of the electromagnetic ($e.m.$) current operator $J^{\mu}$, i.e. to a conserved vector current. Since a very relevant issue is to know to what extent the $e.m.$ properties of a composite system can be understood in terms of the $e.m.$ properties of its constituents, the one-body approximation for the current $J^{\mu}$ has been widely considered. In case of spin-$1/2$ constituents, which is of interest in this work, one has
 \be
      J^{\mu} \simeq J_1^{\mu} = \sum_{j = 1}^N \left[ f_1^{(j)}(q^2) 
      \gamma^{\mu} + f_2^{(j)}(q^2) {i \sigma^{\mu \nu} q_{\nu} \over 2m_j} 
      \right]
      \label{eq:one-body}
 \ee
where $f_{1(2)}^{(j)}(q^2)$ are Dirac ({Pauli) form factors of the $j$-th constituent with mass $m_j$. While the exact current $J^{\mu}$ is covariant with respect to the (interaction-dependent) transverse rotations, the approximate current $J_1^{\mu}$ is not, and a direct manifestation of the loss of the rotational covariance is the so-called angular condition. In this respect let us remind that the form factors appearing in the covariant decomposition of a conserved current can be expressed in terms of the matrix elements of only one component of the current, namely the {\em plus} component (see Ref. \cite{LF}). It may occur however that the number of form factors is less than the number of the independent matrix elements of the {\em plus} component obtained from the application of general properties to the current operator. This means that in such situations a relation among the matrix elements (the angular condition) should occur in order to constrain further their number. The use of the one-body current (\ref{eq:one-body}) may lead to important violations of the angular condition, which can even totally forbid the extraction of the form factors from the matrix elements of the {\em plus} component of the current (see, e.g., Refs. \cite{CAR_rho,Volodia} and \cite{CAR_Delta} in case of the $\rho$-meson and the $N - \Delta(1232)$ transitions, respectively, and also Ref. \cite{CAR_nucleon} for the case of the nucleon elastic form factors).

\indent In Refs. \cite{Frederico_pion}-\cite{Sauer} the Feynman one-loop triangle diagram was calculated in terms of $LF$ variables in a reference frame where $q^+ \ne 0$. As is well known \cite{Zgraph} for point-like bound-state vertexes the triangle diagram is given by the sum of two contributions: the spectator pole and the $Z$-graph. However, for two-fermion systems the one-loop triangle diagram with a point-like bound-state vertex diverges and should be regularized. In Refs. \cite{Frederico_pion}-\cite{Sauer} a simple model of regularizing functions depending only on the momenta of the active constituents was considered. In this case the decomposition of the diagram into the abovementioned two pieces still persists\footnote{Generally speaking the convergence of the one-loop integral is provided by the bound state wave function, which vanishes sufficiently fast at the endpoints. At variance with this general feature, the regularizing function of Refs. \cite{Frederico_pion}-\cite{Sauer} does not suppress at all the endpoint regions.}. The matrix elements for spin-0 and spin-1 bound states for both $(+)$ and $(-)$ components of the $e.m.$ current were analyzed in the limit $q^+ = 0$.

\indent For a spin-0 bound state it was shown \cite{Frederico_pion} that for $q^+ \to 0$ the contribution of the $Z$-graph vanishes for the $(+)$ component of the $e.m.$ current, but it remains finite for the $(-)$ component (the so-called zero longitudinal-momentum mode), providing in this way the full covariance of the $e. m.$ current matrix elements. This means that the same form factor is obtained from the $(+)$ and the $(-)$ components of the current, but the anatomy of the form factor is different in the two cases: from the $(+)$ component the form factor contains only one-body effects, whereas from the $(-)$ component it is the sum of one-body and many-body effects (the longitudinal zero mode).

\indent For a spin-1 bound state a non-vanishing contribution of the $Z$-graph both to the $(-)$ and $(+)$ components of the $e.m.$ current amplitudes was obtained in the limit $q^+ = 0$ \cite{Frederico_rho,Sauer}. Therefore it was concluded that the form factors of the spin-1 bound state are always affected by the $Z$-graph. Since within the $LF$ formalism the $Z$-graph is a many-body process, the results of Refs. \cite{Frederico_rho,Sauer} would imply that a $LF$ one-body approximation consistent with the Feynman triangle diagram cannot be formulated  in any reference frame.

\indent The aim of this work is to analyze the connection between the Feynman triangle diagram and the light-front formalism for spin-0 and spin-1 two-fermion systems. We will limit ourselves to investigate under which conditions the Feynmann triangle diagram can be exactly matched by the so-called Hamiltonian $LF$ formalism, characterized by a fixed number of on-mass-shell constituents. We will not address quantum gauge theories, although we want to stress that part of our analysis might be important also for calculations within such theories (see below, for instance, the impact of the choice of the null-plane on the general description of bound states)\footnote{We thank Prof. R. Jackiw for drawing our attention to this point.}. 

\indent We shall demonstrate that the conclusion of Refs. \cite{Frederico_rho,Sauer} on the effect of the zero-mode upon the form factors strongly relies on the procedure of extracting the form factors from the current amplitudes. We shall formulate a different procedure leading to form factors which represent a fully one-body effect both for spin-0 and spin-1 bound states in the limit $q^+ = 0$. To this end we analyze the components of the $e.m.$ current for the spin-0 system, viz.
 \be
       J^\mu(P_1, P_2) = \langle P_2 | J^\mu | P_1 \rangle ~, 
      \label{eq:amplitude_0}
 \ee
while for spin-1 systems we consider the components of the tensor ${\cal{J}}^{\mu, \alpha \beta}(P_1, P_2)$ related to the $e.m.$ current components as follows
 \be
       \langle P_2, s_2 | J^\mu | P_1, s_1 \rangle = e_{2 \alpha}^*(P_2, 
       s_2) ~ {\cal{J}}^{\mu, \alpha \beta}(P_1, P_2) ~ e_{1 \beta}(P_1, 
       s_1)
       \label{eq:amplitude_1}
 \ee
where $e(P, s)$ is the $LF$ polarization four-vector of the spin-1 system corresponding to spin projection $s$ and total four-momentum $P$.

\indent We study the general structure of the amplitudes for spin-0 and spin-1 bound state along the lines of the covariant approach of Refs. \cite{Karmanov,Carbonell}. The bound state is described by a vertex (or wave function) defined on a hyperplane specified by the light-like four-vector $\omega$, such that $\omega^2 = 0$. As a result, the bound-state wave function acquires an explicit $\omega$-dependence. The interaction of such a bound state with the $e.m.$ field will be conveniently considered in the limit $\omega \cdot q = 0$. The Feynman propagators of the constituents should be also rewritten in a form which explicitly contains the $\omega$-dependence and is similar to the usual $LF$ representation of the Feynman propagator.

\indent Any dependence on the four-vector $\omega$ should disappear if the full amplitude of the bound state coupling with the $e.m.$ current is considered. However, the cancellation of the $\omega$-dependence does not occur if some approximation for the $e.m.$ current [like, e.g., the one-body approximation (\ref{eq:one-body})] is adopted; in other words if one considers the triangle-diagram approximation. The one-loop Feynman amplitude is explicitly Lorentz-covariant but it gets a dependence on the four-vector $\omega$.

\indent The presence of the four-vector $\omega$ leads to the appearance of the additional Lorentz structures in the triangle amplitude and to additional form factors. The physical form factors correspond to Lorentz structures containing only the physical four-vectors $P_1$, $P_2$ and $q = P_2-P_1$, while the form factors entering the Lorentz structures containing $\omega$ are unphysical. All the form factors become functions not only of the squared four-momentum transfer $q^2$, but also of the scalar $\omega \cdot q$. The latter dependence disappears however in the limit $\omega \cdot q = 0$.

The main results we present in this work are as follows:

\begin{itemize}

\item{We give a new definition of the good amplitudes to be used for extracting the physical form factors; more precisely, we determine those amplitudes (hereafter referred to as the {\em good} amplitudes) which contain only physical form factors.

The standard light-front formalism corresponds to a particular choice of the four-vector $\omega$ in the form $\omega = (\omega^-, \omega^+, \vec{\omega}_{\perp}) = (2, 0, \vec{0}_{\perp})$. Let us remind that the standard $LF$ formalism ignores the dependence of the amplitude on the four-vector $\omega$ and this
leads to the loss of the rotational covariance of the amplitudes.

For the reference frame specified by the relations $q^+ = 0$ and $q_x = q_{\perp}$, $q_y = 0$ in the transverse $(x,y)$-plane, the good amplitudes correspond to $\mu = +$ and $\mu = y$ for both spin-0 and spin-1 systems. For spin-1 system one should in addition use $\alpha, \beta \neq (-)$ for the components of the tensor ${\cal{J}}^{\mu, \alpha \beta}$. The good amplitudes defined in this way contain only physical form factors\footnote{Note that our definition of good amplitudes is different from the good component approach of Ref. \cite{FFS}. Moreover, we want to remind that the use of the $\mu = y$ component was firstly proposed in Ref. \cite{Glazek}.}.

According to this definition, for spin-1 systems the $(+)$ component of the e.m. current (\ref{eq:amplitude_1}) used in Refs. \cite{Frederico_rho,Sauer} is not a good amplitude, because it contains the admixture of the bad amplitudes of the tensor ${\cal{J}}^{\mu, \alpha\beta}$ having $\alpha, \beta = -$ after contraction with the longitudinal polarization vectors ($s_1 = 0$ and/or $s_2 = 0$).}

\item{We show that by means of a specific choice of the off-shell behavior of the bound-state vertex the Feynman triangle diagram and the one-body form factors of the standard $LF$ formalism match exactly each other.

To this end we consider an explicit model for the $\omega$-dependent bound-state wave function for {\em off-shell} constituents. We show that it is possible to make a choice of the wave function which takes into account the suppression of the endpoint regions (in accordance with the general properties of the wave functions, valid also in $QCD$), and which guarantees at the same time the following features of the triangle-diagram approximation:

- the appearance of the spectator pole only;

- the suppression of any off-shell effect of the active constituents (in particular, of their instantaneous propagation).

The {\em on-shell} part of the bound-state vertex corresponds to the usual $LF$ wave function and this makes it possible to establish a very useful link with potential models. We stress that such a feature is very important for phenomenological applications, particularly in case of quark models of the hadron structure.

Thus, a physically meaningful one-body approximation, in which all the constituents are on their mass-shells, can be consistently formulated in the limit $q^+ = 0$.}

\end{itemize}

\indent The plan of the paper is as follows. Section 2 and 3 will be devoted to the cases of spin-0 and spin-1 systems, separately, while in Section 4 our covariant $LF$ approach is applied to the evaluation of the $\rho$-meson elastic form factors for comparison with non-covariant $LF$ calculations available in the literature \cite{CAR_rho}. Our main conclusions will be summarized in Section 5.

\section{Spin-0 two-fermion systems}

\indent In this Section we address the calculation of the Feynman triangle diagram in case of a spin-0 system consisting of two spin-$1/2$ fermions. Such an issue has been widely investigated in the literature, particularly in case of pseudoscalar ($PS$) mesons  (see, e.g., Refs. \cite{Frederico_pion,Zgraph,Jaus,pion,Grach,Bmeson}). Thus, hereafter we will consider the triangle diagram contribution ${\cal{J}}_F^{\mu}$ to a $PS \to PS$ transition, viz.
 \be
      {\cal{J}}_F^{\mu}(P_1, P_2) & = & {i \over (2 \pi)^4} \int d^4p ~ 
      \Lambda_1(p, P_1) \Lambda_2(p, P_2) ~ \mbox{Tr}\left\{ {-\dslash{p} + 
      m \over p^2 - m^2 + i \varepsilon} \gamma^5 {\dslash{p}_2 + m_I \over  
      p_2^2 - m_I^2 + i \varepsilon} \Gamma^{\mu} \right. \nonumber \\
      & \cdot & \left. {\dslash{p}_1 + m_I \over p_1^2 - m_I^2 + i 
      \varepsilon} \gamma^5 \right\}
      \label{eq:triangle_0}
 \ee
where $\Gamma^{\mu}$ is the transition current vertex (i.e., $\Gamma^{\mu} = \gamma^{\mu}$ or $\Gamma^{\mu} = i \sigma^{\mu \nu} q_{\nu}$), $P_1$ and $P_2$ are the initial and final momenta of the $PS$ system with mass $M_1$ and $M_2$, $p$ is the momentum of the spectator constituent with mass $m$, $p_1 = P_1 - p$ and $p_2 = P_2 - p$ are the momenta of the active constituent with mass $m_I$. In what follows we will consider the general case $M_2 \neq M_1$, but we will limit ourselves to the case of a conserved current. Finally, $\Lambda_1$ and $\Lambda_2$ are functions of  the constituent momenta describing bound-state vertexes. All the notations are depicted in Fig. 1, where the four-momentum transfer $q$ is given by $q \equiv P_2 - P_1 = p_2 - p_1$. Note that in Eq. (\ref{eq:triangle_0}) the constituent momenta are in general off-mass-shell (i.e., $p_1^2 \neq m_I^2$, $p_2^2 \neq m_I^2$ and $p^2 \neq m^2$), while the system momenta $P_1$ and $P_2$ are on-mass-shell (i.e., $P_1^2 = M_1^2$ and $P_2^2 = M_2^2$).

\begin{figure}[htb]

\vspace{0.25cm}

\centerline{\epsfxsize=10cm \epsfig{file=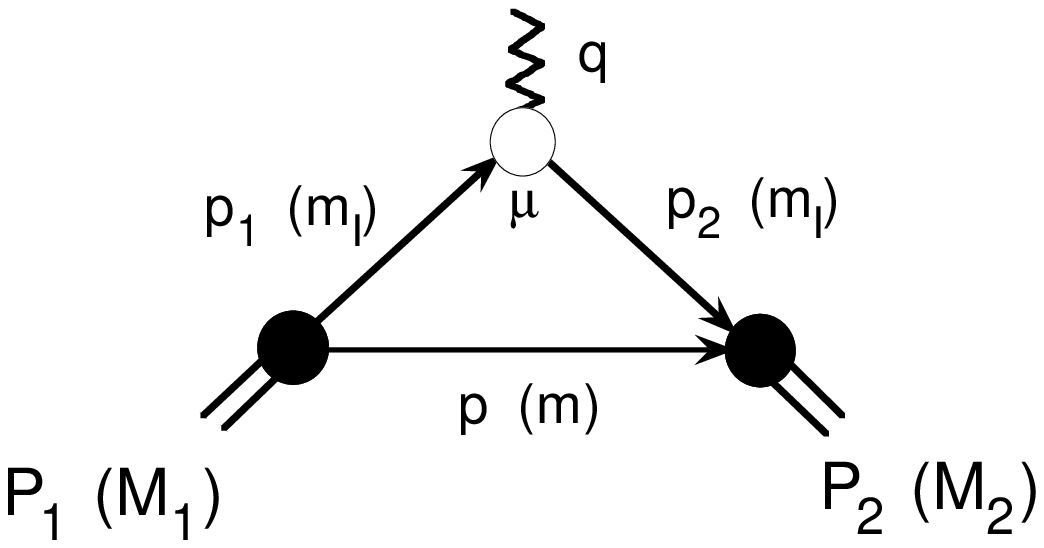}}

{\small \noindent Figure 1. Notations used for the Feynman triangle diagram in case of a spin-0 two-fermion systems.}

\vspace{0.25cm}

\end{figure}

\indent The general decomposition of ${\cal{J}}_F^{\mu}(P_1, P_2)$ into covariant structures is given by
 \be
       {\cal{J}}_F^{\mu}(P_1, P_2) & = & \left[ (P_1 + P_2)^{\mu} - q^{\mu} 
       {M_2^2 - M_1^2 \over q^2} \right] F(q^2) + q^{\mu} (M_2^2 - M_1^2) 
       H(q^2)
       \label{eq:FH}
 \ee
where $F(q^2)$ is the physical form factor describing the $PS \to PS$ transition [i.e., the one appearing in the gauge-invariant term in the r.h.s of Eq. (\ref{eq:FH})], while the term containing the form factor $H(q^2)$ describes possible loss of gauge invariance in the triangle diagram. Note that in case of the elastic process ($M_1 = M_2$) the Lorentz structure proportional to $q^{\mu}$ cannot contribute to the current amplitudes due to time reversal symmetry, and therefore it contains explicitly the factor $(M_2^2 - M_1^2)$.

\indent In terms of $LF$ components (i.e., $p = (p^-, p^+, \vec{p}_{\perp})$, where $p^{\pm} = p^0 \pm p^z$) one gets
 \be
       {\cal{J}}_F^{\mu}(P_1, P_2) & = & {i \over 2 (2 \pi)^4} \int dp^- 
       dp^+ d\vec{p}_{\perp} \mbox{Tr}\left\{ (\dslash{p} + m) (\dslash{P}_2 
       - \dslash{p} + m_I)  \Gamma^{\mu} (\dslash{P}_1 - \dslash{p} + m_I) 
       \right\} \nonumber \\
       & \cdot & {\Lambda_1 \Lambda_2 \over p^+ (P_1^+ - p^+) (P_2^+  - p^+) 
       [p^- - p_{sp}^-] [p^- - p_Z^-] [p^- - p_{Z'}^-]}
      \label{eq:spin0}
 \ee
where the three poles in $p^-$ are explicitly given by
  \be
         p_{sp}^- & = & {m^2 + p_{\perp}^2 \over p^+} - { i \varepsilon 
        \over p^+} ~, \nonumber \\
         p_Z^- & = & P_2^- - {m_I^2 + (\vec{P}_{2 \perp} - 
        \vec{p}_{\perp})^2 \over P_2^+ - p^+} + {i \varepsilon \over P_2^+ - 
        p^+} ~, \nonumber \\
         p_{Z'}^- & = & P_1^- - {m_I^2 + (\vec{P}_{1 \perp} - 
        \vec{p}_{\perp})^2 \over P_1^+ - p^+} + {i \varepsilon \over P_1^+ - 
        p^+}~ .
       \label{eq:poles}
 \ee
Let us now consider the pole structure of Eq. (\ref{eq:spin0}) due only to the constituent propagators. As it is well known (cf. Ref. \cite{Zgraph}), provided the integration over $p^-$ is made convergent by the regularizing functions $\Lambda_1$ and $\Lambda_2$, and applying the Cauchy theorem, four different cases should be analyzed: $p^+ < 0$, $0 \leq p^+ \leq P_1^+$, $P_1^+ \leq p^+ \leq P_2^+$ and $p^+ > P_2^+$, where we have assumed $q^+ \geq 0$ (i.e., $P_2^+ \geq P_1^+$). The first and fourth cases do not contribute to the integral over $p^-$, because the three poles in Eq. (\ref{eq:poles}) have imaginary parts with the same sign. It can be easily seen that the only surviving contributions come from the regions $0 \leq p^+ \leq P_1^+$ and $P_1^+ \leq p^+ \leq P_2^+$. In the former the integration over $p^-$ can be done in the lower half-plane, so that only the pole at $p^- = p_{sp}^-$ (the spectator-pole) contributes. In the latter region the closure in the upper half-plane picks up only the pole at $p^- = p_Z^-$, which yields the so-called $Z$-graph term. Therefore, the triangle diagram can be split into two terms, providing respectively the spectator-pole and the $Z$-graph contributions, namely
 \be
       {\cal{J}}_F^{\mu}(P_1, P_2) = {\cal{J}}_{sp}^{\mu}(P_1, P_2) + 
       {\cal{J}}_{Z}^{\mu}(P_1, P_2)
       \label{eq:sp+Z}
 \ee
where the spectator particle is on-mass-shell in the spectator-pole term (i.e., $p^2 = m^2$ due to the pole $p^- = p_{sp}^- = (m^2 + p_{\perp}^2) / p^+$), while the final active particle is on-mass-shell in the $Z$-graph term (i.e., $p_2^2 = m_I^2$). For an explicit expressions of ${\cal{J}}_{sp}^{\mu}(P_1, P_2) $ and ${\cal{J}}_{Z}^{\mu}(P_1, P_2)$ see, e.g., Ref. \cite{Grach}.

\indent As far as the regularizing functions $\Lambda_1$ and $\Lambda_2$ are concerned, we limit ourselves to  discuss the situation when no further poles besides the ones related to the constituent propagators (see Eq. (\ref{eq:poles})) appear. To this end $\Lambda_1$ and $\Lambda_2$ should be independent on $p^-$. In order to construct such functions in an explicitly covariant way, let us introduce a (null) four-vector $\omega$ which defines the direction normal to the null-plane; the "plus" component of a generic four-vector $k$ is then given by $k^+ = \omega \cdot k$. Let us further define new constituent momenta as ($i=1,2$)
 \be
      \tilde{p}_i & = & p_i - \omega ~ {p_i^2 - m_I^2 \over 2 \omega \cdot 
      p_i} \nonumber \\
      \tilde{p} & = & p - \omega ~ {p^2 - m^2 \over 2 \omega \cdot p}
      \label{eq:ptilde}
 \ee
In this way one has  $\tilde{p}_i^2 = m_I^2$ and $\tilde{p}^2 = m^2$, so that the new momenta coincide with the Feynman ones $p_i$ and $p$ only in the plus and transverse components, while their minus components are constrained by the on-mass-shell conditions, namely: 
 \be
       \tilde{p}_i = (\tilde{p}_i^-, \tilde{p}_i^+, \vec{\tilde{p}}_{i 
       \perp}) = \left( {m_I^2 + p_{i \perp}^2 \over p_i^+}, p_i^+, 
       \vec{p}_{i \perp} \right)
       \label{eq:p_on}
 \ee
and analogously in case of $\tilde{p}$. Note that the momenta $\tilde{p}_i$ and $\tilde{p}$ coincide with those naturally employed in the $LF$ formalism \cite{LF}, since in the latter the constituents are always on their mass-shells. Thus, hereafter they will be referred to as the $LF$ momenta. The regularizing functions $\Lambda_i$ can now be assumed to depend on the scalar $M_{i0} \equiv \sqrt{(\tilde{p_i} + \tilde{p})^2}$, which is clearly independent on $p^-$. More precisely, in order to establish a very useful connection with potential models (see, e.g., \cite{Jaus,pion,Grach,Bmeson}), we assume $\Lambda_i = \Lambda_i(p, P_i, \omega) = \Lambda_i(M_{i0})$ and
 \be
       {1 \over \pi} {\Lambda_i(M_{i0}) \over M_{i0}^2 - M_i^2} \to {1 \over 
       \sqrt{2}} { \sqrt{M_{i0} [1 - (m_I^2 - m^2)^2 / M_{i0}^4]} \over 
       \sqrt{M_{i0}^2 - (m_I - m)^2}} w_i(k_i)
       \label{eq:radial}
 \ee
where $w_i(k)$ is the ($S$-wave) radial function of the $PS$ system, normalized as $\int_0^{\infty} dk ~ k^2 |w_i(k)|^2 = 1$, and $k_i \equiv \sqrt{(M_{i0}^2 + m_I^2 - m^2)^2 - 4 m_I^2 M_{i0}^2} / (2 M_{i0})$.

\indent An important point is that with our regularizing functions (\ref{eq:radial}) suppress the endpoints $p_i^+=0$ and $p_i^+ = P_i^+$, because $w_i(k)$ is a squared integrable function. Therefore, when the limit $q^+ = \omega \cdot q = 0$ is considered, {\em the Z-graph vanishes identically for all current components}. The price to be paid is that now the triangle diagram depends on the four-vector $\omega$ and the covariant decomposition (\ref{eq:FH}) should be modified accordingly. Instead of Eqs.  (\ref{eq:sp+Z}) and (\ref{eq:FH}), respectively, one gets
 \be
       {\cal{J}}_F^{\mu}(P_1, P_2, \omega) \to_{q^+ = 0} 
       {\cal{J}}_{sp}^{\mu}(P_1, P_2, \omega) 
      \label{eq:sp}
 \ee
with
 \be
       {\cal{J}}_{sp}^{\mu}(P_1, P_2, \omega) = I^{\mu}(P_1, P_2) + 
       B^{\mu}(P_1, P_2, \omega)
       \label{eq:sp_0}
 \ee
where $I^{\mu}(P_1, P_2)$ represents all the covariant structures depending only on the physical four-vectors $P_i$, while $B^{\mu}(P_1, P_2, \omega)$ contains all possible $\omega$-dependent structures. Explicitly, for spin-0 systems one has
 \be
       I^{\mu}(P_1, P_2) = \left[ (P_1 + P_2)^{\mu} - q^{\mu} {M_2^2 - M_1^2 
       \over q^2} \right] F(q^2) + q^{\mu} (M_2^2 - M_1^2)  H(q^2)
       \label{eq:good_0}
 \ee
and
 \be
        B^{\mu}(P_1, P_2, \omega) = {\omega^{\mu} \over \omega \cdot P_1} 
        B(q^2)
       \label{eq:B_0}
 \ee
where $B(q^2)$ is the unphysical form factor describing the spurious dependence of the current matrix elements upon the four-vector $\omega$ in case of approximate current operators (cf. Refs. \cite{Karmanov,Carbonell}). If besides the triangle diagram all higher-order diagrams were included, the full result should be independent on $\omega$, i.e., $B(q^2) = 0$ leading to $B^{\mu} = 0$.  From Eqs. (\ref{eq:sp_0}-\ref{eq:B_0})  it follows that the current component $\mu = -$ is the only one to be affected by the spurious form factor $B(q^2)$. 

\indent In Eq. (\ref{eq:B_0}) the four-vector $\omega$ appear in the combination $\omega^{\mu} / (\omega \cdot P_1)$ because the amplitudes should not change if $\omega$ is multiplied by a c-number. In this respect note that the appropriate combinations $\omega^{\mu} / (\omega \cdot p_i)$ and $\omega^{\mu} / (\omega \cdot p)$ appear in Eq. (\ref{eq:ptilde}). 

\indent An important point is that the form factors $F(q^2)$, $H(q^2)$ and $B(q^2)$ may be functions not only of the squared momentum transfer $q ^2$, but also on the scalar $(\omega \cdot q) / (\omega \cdot P_1)$. The latter dependence disappears however in the limit $q^+ = \omega \cdot q = 0$. Moreover, notice that the four-vector $\tilde{q} \equiv \tilde{p}_2 - \tilde{p}_1$ differs from the momentum transfer $q$ in its minus component. However, we have always $\tilde{q}^2 = q^2$ because of $q^+ = 0$. This a very welcome and physically meaningful feature since in the triangle diagram the squared four-momentum transfer seen by the constituents is always $q^2$. 

\indent We want now to understand when the triangle diagram (\ref{eq:sp_0}) and the $LF$ formalism match each other, i.e. under which conditions (and for which components of the $e.m.$ current) all the particles can be put on their mass-shells in the triangle diagram. To this end let us first consider the trace appearing in the r.h.s of Eq. (\ref{eq:spin0}) for $p^- = p_{sp}^- = \tilde{p}^-$ (spectator pole). Using standard trace theorems and starting with the case $\Gamma^{\mu} = \gamma^{\mu}$, one finds
 \be
       {1 \over 2} \mbox{Tr}\{ ... \gamma^{\mu} ... \} & = & [(\tilde{p} + 
       \tilde{p}_2)^2 - (m - m_I)^2] p_1^{\mu} + [ (\tilde{p} + 
       \tilde{p}_1)^2 - (m - m_I)^2]  p_2^{\mu}  \nonumber \\
      & + & [(\tilde{p}_1 - \tilde{p}_2)^2]  p^{\mu} \nonumber \\
      & + & [p_1^- - \tilde{p}_1^-] (p^+ p_2^{\mu} -  p^{\mu} p_2^+) + 
      [p_2^- - \tilde{p}_2^-] (p^+ p_1^{\mu} - p^{\mu} p_1^+) ~, 
      \label{eq:trace0}
 \ee
which has to be compared with the corresponding $LF$ trace given by (cf., e.g., Ref. \cite{Jaus})
 \be
       {1 \over 2} \mbox{Tr}\{ ... \gamma^{\mu} ... \}_{LF} & = & 
       [(\tilde{p} +  \tilde{p}_2)^2 - (m - m_I)^2] \tilde{p}_1^{\mu} + 
       [(\tilde{p} + \tilde{p}_1)^2 - (m - m_I)^2]  \tilde{p}_2^{\mu}  
      \nonumber \\
      & + & [(\tilde{p}_1 - \tilde{p}_2)^2]  \tilde{p}^{\mu} ~.
      \label{eq:trace0_LF}
 \ee
Since the mismatch between the $LF$ and the Feynman momenta is limited only to the minus components, it is clear that at least one should have $\mu \neq -$ in Eqs. (\ref{eq:trace0}-\ref{eq:trace0_LF}). An important feature of Eq. (\ref{eq:trace0}) is that the choice $\mu = +$ guarantees that the terms proportional to $(p_i^- - \tilde{p}_i^-)$ [$i = 1, 2$] are identically vanishing, so that the traces (\ref{eq:trace0}) and (\ref{eq:trace0_LF}) coincide. We point out that the terms proportional to $(p_i^- - \tilde{p}_i^-)$ are the contributions of the off-mass-shell effects generated by the mismatch between the numerators of the Feynman propagator $(\dslash{p}_i + m_I)$ and of the $LF$ propagator $(\dslash{\tilde{p}}_i + m_I)$ of the active particle, viz. 
 \be
       (\dslash{p}_i + m_I) -  (\dslash{\tilde{p}}_i + m_I) = \gamma^+ 
       {p_i^- - \tilde{p}_i^- \over 2} = {\dslash{\omega} \over 2 \omega 
       \cdot p_i} (p_i^2 - m_I^2)
      \label{eq:IP}
 \ee
representing the so-called instantaneous propagation term (absent for on-mass-shell particles). Thus, the $IP$ terms of the active constituent do not contribute to the trace (\ref{eq:trace0}) when $\mu = +$ and therefore the triangle diagram and the $LF$ formalism match exactly for $\mu = +$, provided $q^+ = 0$.

\indent In order to complete the calculation of the triangle diagram we have to evaluate the quantities $(p_{sp}^- - p_Z^-)$ and $(p_{sp}^- - p_{Z'}^-)$, which appear in the spectator-pole term; one has
 \be
        p_{sp}^- - p_Z^- & = & {1 \over P_2^+} \left[ (\tilde{p} + 
        \tilde{p}_2)^2 - M_2^2 \right] \equiv {1 \over P_2^+} \left[ 
        M_{20}^2 - M_2^2 \right] ~, \nonumber \\
        p_{sp}^- - p_{Z'}^- & = & {1 \over P_1^+} \left[ (\tilde{p} + 
        \tilde{p}_1)^2 - M_1^2 \right]  \equiv {1 \over P_1^+} \left[ 
        M_{10}^2 - M_1^2 \right]
       \label{eq:spectator}
 \ee
where $M_{10}$ and $M_{20}$ represent the free mass of the initial and final system, respectively. In terms of intrinsic $LF$ variables, defined as
 \be
       \xi & \equiv & {p_1^+ \over P_1^+} = \left( 1 - {p^+ \over P_1^+} 
       \right) ~, \nonumber \\
       \vec{k}_{\perp} & = & \vec{p}_{1 \perp} - \xi \vec{P}_{1 \perp} = 
       - \vec{p}_{\perp} + (1 - \xi) \vec{P}_{1 \perp} ~, 
       \label{eq:LFvariables}
 \ee
one gets
 \be
       M_{10}^2 = {m_I^2 + k_{\perp}^2 \over \xi} + {m^2 + k_{\perp}^2 \over 
       1 - \xi} ~, \nonumber \\
       M_{20}^2 = {m_I^2 + {k'}_{\perp}^2 \over \xi} + {m^2 + {k'}_{\perp}^2 
       \over 1 - \xi}
       \label{eq:Mi0}
 \ee
with $\vec{k'}_{\perp} = \vec{k}_{\perp} + (1 - \xi) \vec{q}_{\perp}$ in the limit $q^+ = 0$. Thus, we have
 \be
       {\cal{J}}_F^+(P_1, P_2, \omega) \to_{q^+ = 0} I^+(P_1, P_2) = 2 P_1^+ 
       ~ F(q^2)
      \label{eq:Jplus}
 \ee
with (for $\Gamma^+ =  \gamma^+$)
 \be
       F(q^2) & = & {1 \over 2 (2 \pi)^3} \int_0^1 d\xi \int 
       d\vec{k}_{\perp} {\Lambda_1 \over M_{10}^2 - M_1^2} {\Lambda_2 \over 
       M_{20}^2 - M_2^2} \nonumber \\
       & \cdot & {[M_{10}^2 + M_{20}^2 - 2 (m - m_I)^2] \xi + q^2 (1 - \xi) 
       \over \xi^2 (1 - \xi)} ~.
      \label{eq:F0}
 \ee
Using Eq. (\ref{eq:radial}) and putting $\mu(\xi) \equiv m_I ~ (1 - \xi) + m ~ \xi$, one gets
 \be
       F(q^2) & = & {1 \over 4 \pi} \int d\xi \int d\vec{k}_{\perp} ~ 
       \sqrt{A_1(\xi, \vec{k}_{\perp}) ~ A_2(\xi, \vec{k'}_{\perp})} ~ 
       w_1(k^2) w_2({k'}^2) \nonumber \\
       & \cdot & {\mu^2(\xi) + \vec{k}_{\perp} \cdot \vec{k'}_{\perp} \over 
       \sqrt{\mu^2(\xi) + k_{\perp}^2} \sqrt{\mu^2(\xi) + {k'}_{\perp}^2}}
       \label{eq:F0_LF}
 \ee
where $A_1(\xi, \vec{k}_{\perp}) = M_{10} [ 1 - (m_I^2 - m^2)^2 / M_{10}^4] / 4 \xi (1 - \xi)$ is a normalization factor, with $k^2 \equiv k_{\perp}^2 + k_n^2$ and $k_n \equiv M_{10} (\xi - 1/2) + (m^2 - m_I^2) / 2M_{10}^2$. Eq. (\ref{eq:F0_LF}) coincides with the result obtained within the $LF$ formalism in, e.g., Refs. \cite{Jaus,pion,Grach,Bmeson}. For completeness, let us briefly remind the general structure of the $LF$ result, which at $q^+ = 0$ explicitly reads as
 \be
       {\cal{J}}_{LF}^+(P_1, P_2) = 2P_1^+ ~ _{LF}\langle PS_2 | \bar{u}_2 
       \Gamma^+ u_1 | PS_1\rangle_{LF}
       \label{eq:LFresult_0}
 \ee
where $u_1$ and $u_2$ are Dirac spinors and
 \be
        | PS_1\rangle_{LF} = \mbox{R}^{(0)}(\xi, \vec{k}_{\perp}) w_1(k) 
        \sqrt{{A_1(\xi, \vec{k}_{\perp}) \over 4\pi}}
        \label{eq:LFwf_0}
 \ee
with $\mbox{R}^{(0)}(\xi, \vec{k}_{\perp})$ being the product of (generalized) Melosh rotation spin matrices \cite{Melosh} appropriate for a spin-0 system. Note that in terms of Dirac spinors the spin matrix $\mbox{R}^{(0)}(\xi, \vec{k}_{\perp})$ can be written as  (cf., e.g., Ref. \cite{Jaus})
 \be
       \left[ \mbox{R}^{(0)}(\xi, \vec{k}_{\perp}) \right]_{\lambda_1 
       \lambda} = {1 \over \sqrt{2}} {1 \over \sqrt{M_{10}^2 - (m_I - m)^2}} 
       \bar{u}(\tilde{p}_1, \lambda_1) \gamma^5 v(\tilde{p}, \lambda) ~,
       \label{eq:Melosh_0}
 \ee
so that the $LF$ trace (\ref{eq:trace0_LF}) can be rewritten as
 \be
       \mbox{Tr}\{ ... \Gamma^{\mu} ... \}_{LF} \propto \sum_{\lambda_1 
       \lambda_2 \lambda} \left[ \mbox{R}^{(0)}(\xi, \vec{k'}_{\perp}) 
       \right]_{\lambda \lambda_2}^* ~ \overline{u}(\tilde{p}_2, \lambda_2) 
       \Gamma^{\mu} u(\tilde{p}_1, \lambda_1) ~ \left[ \mbox{R}^{(0)}(\xi, 
       \vec{k}_{\perp}) \right]_{\lambda_1 \lambda} ~,
       \label{eq:Melosh_LF}
 \ee
leading to ${\cal{J}}_{LF}^+(P_1, P_2) = I^+(P_1, P_2)$.

\indent The case of the transverse components of the current ($\mu = \perp$) deserves a closer look. Let us consider the limit $q^+ = 0$ and choose the direction of the $x$-axis along $\vec{q}_{\perp}$, so that the $y$-axis is the direction orthogonal to $\vec{q}_{\perp}$ in the transverse plane (i.e., $q_y = 0$ by definition). For $\mu = y$ the $IP$ terms proportional to $(p_i^- - \tilde{p}_i^-)$ in Eq. (\ref{eq:trace0}), are {\em odd} in the integration variable $k_y$, which is unaffected by the virtual photon absorption process. Thus, after integration over $k_y$ the $IP$ effects of the active constituent in the trace (\ref{eq:trace0}) are identically vanishing for $\mu = y$, provided the regularizing functions $\Lambda_i$ are even in $k_y$, as it is the case of Eq. (\ref{eq:radial}). The same situation does not occur when $\mu = x$, since now both the trace (\ref{eq:trace0}) and the product $\Lambda_1 \Lambda_2$ are neither even nor odd in $k_x$ because ${k'}_x = k_x + (1 - \xi) q_x$. Moreover, according to the decomposition (\ref{eq:sp_0}-\ref{eq:B_0}) the $\mu = x$ component is affected by the form factor $H(q^2)$ related to the loss of gauge invariance of the triangle diagram, while both the $\mu = +$ and $\mu = y$ components are independent on $H(q^2)$ \footnote{As far as the elastic process is considered,  after shifting the integration variable from $k_x$ to $K_x \equiv (k_x + {k'}_x) / 2$ one has that the product $\Lambda_1 \Lambda_2$ is even in $K_x$. Since on the contrary the full trace (\ref{eq:trace0}) is odd in $K_x$, the component ${\cal{J}}^{x}$ is identically vanishing, as it is required by time reversal symmetry.}.

\indent Thus, for spin-0 systems the triangle diagram and the $LF$ formalism match exactly for $\mu = +$ and $\mu = y$, provided the limit $q^+ = 0$ is considered. It is straightforward to check that from Eqs. (\ref{eq:trace0}) and (\ref{eq:LFvariables}) the use of the $(y)$ and the $(+)$ components of the current leads to the extraction of the same physical form factor $F(q^2)$, given explicitly by Eq. (\ref{eq:F0_LF}).

\indent For completeness let us now consider the Pauli coupling $\Gamma^{\mu} = i \sigma^{\mu \nu} q_{\nu}$. The trace appearing in Eq. (\ref{eq:spin0}) is now given by
 \be
        {1 \over 4} \mbox{Tr}\{ ... i \sigma^{\mu \nu} q_{\nu} ... \} & = & 
        m [p_1^{\mu} (p_2 \cdot q) - p_2^{\mu} (p_1 \cdot q)] + m_I [p^{\mu} 
        (p_2 \cdot q) - p_2^{\mu} (p \cdot q)] \nonumber \\
        & + & m_I [p_1^{\mu} (p \cdot q) - p^{\mu} (p_1 \cdot q)]
       \label{eq:Pauli}
 \ee
and therefore it is enough to choose $\mu \neq -$ and $q^+ = 0$ to ensure that the trace (\ref{eq:Pauli}) involves only on-mass-shell momenta. In this way, the net effect of the $IP$ terms (\ref{eq:IP}) is identically vanishing in Eq. (\ref{eq:Pauli}). Thus, in case of the Pauli coupling the limit $q^+ = 0$ allows to put all the particles on their mass-shells for $\mu = +, \perp$.

\indent Before closing this Section let us now briefly compare our findings with the results of Ref. \cite{Frederico_pion}. There the regularizing functions $\Lambda_1$ and $\Lambda_2$ are assumed to depend on the momenta of the active constituents, namely:
 \be
       \Lambda_i = \Lambda_i(p_i^2) = \left[ {1 \over p_i^2 - \mu_R^2 +i 
       \varepsilon} \right]^n
      \label{eq:choice_Fre}
 \ee
where $n$ is a positive integer and $\mu_R$ plays the role of a regularizing cutoff. In this way the regularizing functions depend explicitly on $p^-$ and introduce further poles in Eq. (\ref{eq:spin0}). The latter are however located in the upper half-plane for $0 \leq p^+ \leq P_1^+$, so that they do not affect the spectator contribution. Any way, for $P_1^+ \leq p^+ \leq P_2^+$ the poles of Eq. (\ref{eq:choice_Fre}) affects the $Z$-graph contribution. The important difference with our choice (\ref{eq:radial}) is that the regularizing functions (\ref{eq:choice_Fre}) do not introduce any dependence upon the four-vector $\omega$, but they do not suppress at all the endpoints $p_i^+=0$ and $p_i^+ = P_i^+$. Therefore, in case of the current component $\mu = -$ the triangle diagram receives always the contributions of both the spectator-pole and the $Z$-graph even in the limit $q^+ = 0$, as properly pointed out in Ref. \cite{Frederico_pion}. The inclusion of the zero-mode for $\mu = -$ turns out \cite{Frederico_pion} to be essential in order to fulfill the covariance of the triangle diagram.

\indent We want to point out that the zero-mode of Ref. \cite{Frederico_pion} and the spurious form factor $B(q^2)$ appearing in Eq. (\ref{eq:B_0}) are not directly related each other.  However, the physical origin of these terms is the same, i.e. the loss of the rotational covariance in the $LF$ formalism, which manifests itself in a different way according to the specific choice of the form of the bound-state vertexes.

\section{Spin-1 two-fermion systems}

\indent As anticipated in the Introduction, in case of spin-1 systems we consider the triangle diagram amplitudes (see Fig. 2), which read as
 \be
      {\cal{J}}_F^{\mu, \alpha \beta}(P_1, P_2) & = & {i \over (2 \pi)^4} 
      \int d^4p ~ \Lambda_1(p, P_1) \Lambda_2(p, P_2) ~ \mbox{Tr}\left\{ 
      {-\dslash{p} + m \over p^2 - m^2 + i \varepsilon} V^{\alpha} 
      {\dslash{p}_2 + m_I \over p_2^2 - m_I^2 + i \varepsilon} \right. 
      \nonumber \\
      & \cdot & \left. \Gamma^{\mu}  {\dslash{p}_1 + m_I \over p_1^2 - m_I^2 
      + i \varepsilon} V^{\beta} \right\}
      \label{eq:triangle_1}
 \ee
where $V^{\alpha}$ ($V^{\beta}$) describes the Lorentz structure of the final (initial) spin-1 vertex.

\begin{figure}[htb]

\vspace{0.25cm}

\centerline{\epsfxsize=10cm \epsfig{file=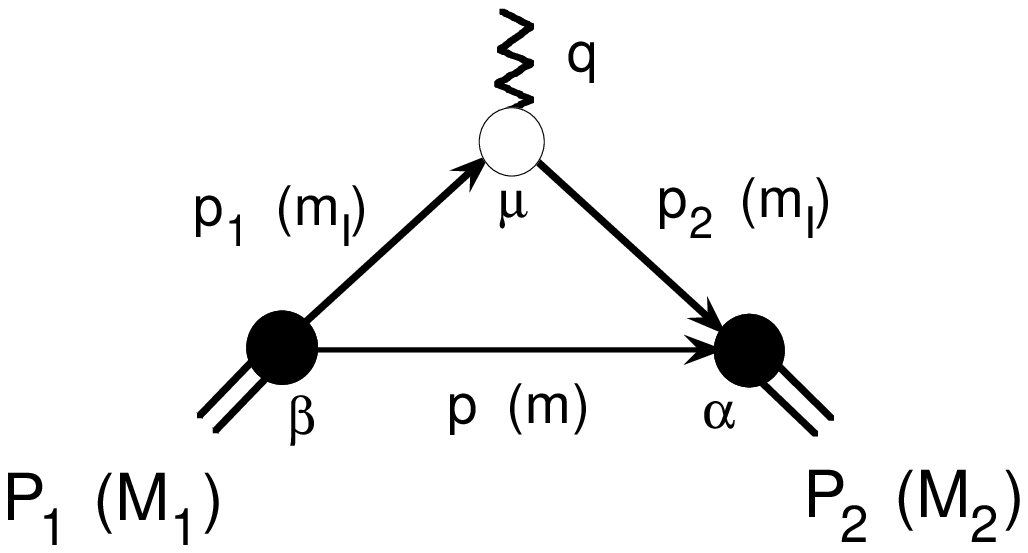}}

{\small \noindent Figure 2. Notations used for the Feynman triangle diagram in case of a spin-1 two-fermion systems.}

\vspace{0.25cm}

\end{figure}

\indent In what follows we will limit ourselves to the case of an $S$-wave internal motion, so that the vertexes $V^{\alpha}$ and $V^{\beta}$ are explicitly given by
 \be 
       V^{\alpha} = \gamma^{\alpha} - {(p_2 - p)^{\alpha} \over M_{20} + 
       m_I + m} ~, ~~~~~~~~~~~~~~~~
       V^{\beta} = \gamma^{\beta} - {(p_1 - p)^{\beta} \over M_{10} + 
       m_I + m} ~.
      \label{eq:vertexes}
 \ee
Let us first consider the Dirac coupling $\Gamma^{\mu} = \gamma^{\mu}$ and for ease of presentation we limit the structures $V^{\alpha}$ and $V^{\beta}$ to $\gamma^{\alpha}$ and $\gamma^{\beta}$, respectively. In case of our regularizing functions (\ref{eq:radial}) {\em any $Z$-graph effect is completely absent in all components of the tensor} (\ref{eq:triangle_1}) when the limit $q^+ = 0$ is considered, namely
 \be
       {\cal{J}}_F^{\mu, \alpha \beta}(P_1, P_2, \omega) \to_{q^+ = 0} 
      {\cal{J}}_{sp}^{\mu, \alpha \beta}(P_1, P_2, \omega) ~.
       \label{eq:sp_tensor}
 \ee
where
 \be
       {\cal{J}}_{sp}^{\mu, \alpha \beta}(P_1, P_2, \omega) & = & {1 \over 2 
       (2 \pi)^3} \int dp^+ d\vec{p}_{\perp} \mbox{Tr}\left\{ (- 
       \dslash{\tilde{p}} + m)  \gamma^{\alpha} (\dslash{P}_2 -  
       \dslash{\tilde{p}} + m_I)  \gamma^{\mu} (\dslash{P}_1 - 
       \dslash{\tilde{p}} + m_I) \gamma^{\beta} \right\} \nonumber \\
       & \cdot & {\Lambda_1(M_{10}) ~ \Lambda_2(M_{20}) \over p^+ (1 - 
       p^+ / P_1^+) (1- p^+ / P_2^+) [M_{10}^2 - M_1^2] [M_{20}^2 - M_2^2]} 
       ~.
      \label{eq:spin1}
 \ee
 The explicit calculation of the trace appearing in the r.h.s. of Eq. (\ref{eq:spin1}) yields
 \be
       {1 \over 2} \mbox{Tr}\{ ... \gamma^{\mu} ...\} & = & (m m_I + 
       \tilde{p} \cdot \tilde{p}_1) (g^{\mu \beta} p_2^{\alpha} - g^{\mu 
       \alpha} p_2^{\beta} + g^{\alpha \beta} p_2^{\mu}) 
       \nonumber \\
       & + & (m m_I + \tilde{p} \cdot \tilde{p}_2)  (g^{\mu \alpha} 
       p_1^{\beta} - g^{\mu \beta} p_1^{\alpha} +  g^{\alpha \beta} 
       p_1^{\mu}) 
       \nonumber \\ 
       & + & (\tilde{p}_1 \cdot \tilde{p}_2 - m_I^2) (g^{\mu \beta} 
        \tilde{p}^{\alpha} - g^{\alpha \beta} \tilde{p}^{\mu} + g^{\mu 
       \alpha} \tilde{p}^{\beta}) 
       \nonumber \\
       & - & p_1^{\mu} (\tilde{p}^{\alpha} p_2^{\beta} + \tilde{p}^{\beta} 
       p_2^{\alpha}) -p_2^{\mu} (\tilde{p}^{\alpha} p_1^{\beta} + 
       \tilde{p}^{\beta} p_1^{\alpha}) - \tilde{p}^{\mu} (p_2^{\alpha} 
       p_1^{\beta} - p_2^{\beta} p_1^{\alpha}) 
       \nonumber \\
       & + & (p_1^- - \tilde{p}_1^-) \{ g^{\alpha \beta} [p^+ p_2^{\mu} - 
       \tilde{p}^{\mu} p_2^+] + g^{\mu \alpha} [\tilde{p}^{\beta} p_2^+ - 
       p^+ p_2^{\beta}] + g^{\mu \beta} [ \tilde{p}^{\alpha} p_2^+ + p^+ 
       p_2^{\alpha}] \} 
       \nonumber \\ 
       & + & (p_2^- - \tilde{p}_2^-) \{ g^{\alpha \beta} [p^+ p_1^{\mu} - 
       \tilde{p}^{\mu} p_1^+] + g^{\mu \alpha} [p^+ p_1^{\beta} + 
       \tilde{p}^{\beta} p_1^+] + g^{\mu \beta} [\tilde{p}^{\alpha} p_1^+ - 
       p^+ p_1^{\alpha}] \} ~. ~~~~~~~
       \label{eq:trace1}
 \ee
Let us now proceed in a way analogous to the case of spin-0 systems discussed in the previous Section. Since the mismatch between the $LF$ and the Feynman momenta is limited only to the $(-)$ components, all the indexes $\alpha$, $\beta$ and $\mu$ in Eq. (\ref{eq:trace1}) should be at least different from $(-)$. Moreover, for $\mu = +$ and $\alpha, \beta \neq -$ the trace (\ref{eq:trace1}) involves only on-mass-shell momenta and the net effect of the terms proportional to $(p_i^- - \tilde{p}_i^-)$, corresponding to the $IP$ contributions of the active constituent, is identically vanishing. However, at variance with the spin-0 case, the $IP$ terms do not vanish in general for $\mu = y$. Now, the key observations are: ~ i) the $IP$ contributions are off-shell effects in the triangle diagram, and ~ ii) the components of the bound-state vertex $V^{\beta(\alpha)}$ with $\beta(\alpha) \neq -$ do not involve the $(-)$ components of the constituent momenta. Therefore, it is natural to ask whether it is possible to redefine the vertex $V^{\beta(\alpha)}$ in such a way to kill the $IP$ terms. This is indeed possible by introducing an $\omega$-dependent term in the bound-state vertexes, namely
 \be
       \overline{V}^{\beta} & = & V^{\beta} + (m_I - \dslash{p}_1) 
       {\dslash{\omega} \over 2(\omega \cdot p_1)} V^{\beta} \nonumber \\
       \overline{V}^{\alpha} & = & V^{\alpha} + V^{\alpha} 
       {\dslash{\omega} \over 2(\omega \cdot p_2)} (m_I - \dslash{p}_2)
       \label{eq:V_new}
 \ee
In this way the $IP$ contributions are exactly canceled out in the tensor (\ref{eq:spin1}), because one has $(m_I + \dslash{p}_1)  ~ \overline{V}^{\beta} = (m_I + \dslash{\tilde{p}}_1) ~ V^{\beta}$ and $\overline{V}^{\alpha} ~ (m_I + \dslash{p}_2) = V^{\alpha} ~ (m_I + \dslash{\tilde{p}}_2)$. At the same time the on-shell part of  bound-state vertexes is unchanged, i.e., one has $\bar{u}(\tilde{p}_{1(2)}, \lambda_{1(2)}) \overline{V}^{\beta(\alpha)} v(\tilde{p}, \lambda) = \bar{u}(\tilde{p}_{1(2)}, \lambda_{1(2)}) V^{\beta(\alpha)} v(\tilde{p}, \lambda)$, where $u$ and $v$ are Dirac spinors. Since $\dslash{\omega} ~ \dslash{\omega} = \omega \cdot \omega = 0$,  the vertex components $\overline{V}^{\beta(\alpha)}$ with $\beta(\alpha) \neq -$ do not involve the $(-)$ components of the constituent momenta. Note that the cancellation of the $IP$ contributions has been realized through the bound-state vertexes (\ref{eq:V_new}) in a fully covariant way, i.e., without modifying the covariance properties of the Feynman triangle diagram\footnote{In the approach of Refs. \cite{Karmanov,Carbonell} the absence of $IP$ terms due to propagators attached to the bound-state vertexes is fulfilled automatically. In case of fermion propagators not attached to the bound-state vertexes the situation might be more complicated, but these cases are not related to the analysis of the triangle diagram presented in this work. We thank V. Karmanov for valuable discussions on this point.}.

Thus, by means of Eqs. (\ref{eq:radial}) and (\ref{eq:V_new}) we have constructed an explicit (covariant) model for spin-1 bound-state wave function which takes into account the suppression of the endpoint regions (in accordance with the general properties of the wave functions, valid also in $QCD$), and which guarantees at the same time the appearance of only the spectator pole in the Feynman triangle diagram and the suppression of any off-shell effect of the active constituents (in particular, of their instantaneous propagation). The {\em on-shell} part of the bound-state vertex corresponds to the usual $LF$ wave function (see next Section) and this makes it possible to establish a very useful link with potential models.

\indent The generalizations of the above findings to the cases of the full $S$-wave structures (\ref{eq:vertexes}) as well as to the Pauli coupling $\Gamma^{\mu} = i \sigma^{\mu \nu} q_{\nu}$ are straightforward and will not be reported here. Note that our results hold as well also in case of $D$-wave internal motion.

\indent Let us now show that the number of {\em  good} components of the tensor (\ref{eq:spin1}) are enough to extract in a unique way the form factors for spin-1 systems. For ease of presentation we limit ourselves to the case of an elastic process. Let us start with the general covariant decomposition of the tensor ${\cal{J}}_{sp}^{\mu, \alpha \beta}(P, P', \omega)$ for $P^2 = {P'}^2 = M^2$, viz.
 \be
       {\cal{J}}_{sp}^{\mu, \alpha \beta}(P, P', \omega) = I^{\mu, \alpha 
       \beta}(P, P') + B^{\mu, \alpha \beta}(P, P', \omega)
       \label{eq:sp_1}
 \ee
where the tensor $I^{\mu, \alpha \beta}(P, P')$ is independent of the four-vector $\omega$, while $B^{\mu, \alpha \beta}(P, P', \omega)$ contains all the possible covariant structures depending on $\omega$. Explicitly, one has
 \be
        I^{\mu, \alpha \beta}(P, P') & = & - (P + P')^{\mu} \left\{ 
        F_1(q^2) \left[ g^{\alpha \beta} - {P^{\alpha} P^{\beta} \over M^2} 
        - {{P'}^{\alpha} {P'}^{\beta} \over M^2} + {{P'}^{\alpha} P^{\beta} 
        \over M^2} {P \cdot P' \over M^2} \right] \right. \nonumber \\
        & + & \left. {F_2(q^2) \over 2 M^2} \left( q^{\alpha} - {P' \cdot q 
        \over M^2} {P'}^{\alpha} \right) \left( q^{\beta} - {P \cdot q \over 
        M^2} P^{\beta} \right) \right\} \nonumber \\
        & + & F_3(q^2) \left\{ \left( g^{\mu \alpha} - {{P'}^{\mu} 
        {P'}^{\alpha} \over M^2} \right) \left( q^{\beta} - {P \cdot q \over 
        M^2} P^{\beta} \right) \right. \nonumber \\
        & - & \left. \left( g^{\mu \beta} - {P^{\mu} P^{\beta} \over M^2} 
        \right) \left( q^{\alpha} - {P' \cdot q \over M^2} {P'}^{\alpha} 
        \right) \right\} \nonumber \\
        & - & (P + P')^{\mu} \left\{ H_1(q^2) {{P'}^{\alpha} P^{\beta} \over 
        M^2} + {H_2(q^2) \over 2 M^2} \left(q^{\alpha} P^{\beta} - q^{\beta} 
        {P'}^{\alpha} \right) \right\} \nonumber \\
        & + & H_3(q^2) \left( g^{\mu \alpha} P^{\beta} + g^{\mu \beta} 
        {P'}^{\alpha} \right) + H_4(q^2) q^{\mu} {q^{\alpha} P^{\beta} + 
        q^{\beta} {P'}^{\alpha} \over M^2}
        \label{eq:F+H}
 \ee
and (cf. Ref. \cite{Karmanov})
 \be
       B^{\mu, \alpha \beta}(P, P', \omega) & = & {M^2 \over 2 (\omega \cdot 
       P)} \omega^{\mu} ~ \left[ B_1(q^2) ~ g^{\alpha \beta} + B_2(q^2) ~ 
       {q^{\alpha} q^{\beta} \over M^2} + M^2 B_3(q^2) ~ {\omega^{\alpha} 
       \omega^{\beta} \over (\omega \cdot P)^2} \right. \nonumber \\
       & + & \left. B_4(q^2) ~ {q^{\alpha} \omega^{\beta} - q^{\beta} 
       \omega^{\alpha} \over 2 (\omega \cdot P)} \right] + (P + P')^{\mu} 
       \left[ M^2 B_5(q^2) ~ {\omega^{\alpha} \omega^{\beta} \over (\omega 
       \cdot P)^2} \right. \nonumber \\
       & + & \left. B_6(q^2) ~ {q^{\alpha} \omega^{\beta} - q^{\beta} 
       \omega^{\alpha} \over 2 (\omega \cdot P)} \right] + M^2 B_7(q^2) ~ 
       {g^{\mu \alpha} \omega^{\beta} + g^{\mu \beta} \omega^{\alpha} \over 
       (\omega \cdot P)} \nonumber \\
       & + & B_8(q^2) ~ q^{\mu} ~ {q^{\alpha} \omega^{\beta} + q^{\beta} 
       \omega^{\alpha} \over 2 (\omega \cdot P)}
       \label{eq:B_1}
 \ee
where all the covariant structures included in Eqs. (\ref{eq:F+H}-\ref{eq:B_1}) satisfy both parity and time reversal symmetries. A direct inspection of Eq. (\ref{eq:B_1}) reveals that for $\alpha, \beta, \mu \neq -$ one has $B^{\mu, \alpha \beta}(P, P', \omega) = 0$. Therefore the good components of the tensor (\ref{eq:spin1}) coincide with the corresponding ones of the tensor $I^{\mu, \alpha \beta}(P, P')$, which are independent on the spurious four-vector $\omega$.

\indent In Eq. (\ref{eq:F+H}) there are seven form factors, namely the three form factors $F_i(q^2)$ ($i = 1, 2, 3$) and the four form factors $H_j(q^2)$ ($j = 1, ..., 4$). The form factors $F_i(q^2)$ appear in covariant structures which are transverse to all the external momenta $P$, $P'$ and $q$, while the form factors $H_j(q^2)$ describe the loss of transversity (including the loss of gauge invariance) in the triangle diagram. Therefore, since $e_1(P, s_1) \cdot P = e_2(P', s_2) \cdot P' = 0$, only the form factors $F_i(q^2)$ [$i = 1, 2, 3$] and $B_k(q^2)$ [$k = 1, 2, ... 8$] appear in the decomposition of the $e.m.$ current matrix elements,  viz.
 \be
       {\cal{J}}_{s_2, s_1}^{\mu}(P, P', \omega) & \equiv & e_{2 
       \alpha}^*(P', s_2) ~ {\cal{J}}_{sp}^{\mu, \alpha \beta}(P,  P', 
       \omega) ~ e_{1 \beta}(P, s_1) = - (P + P')^{\mu} \nonumber \\
       & \cdot & \left\{ F_1(q^2) e_2^*(P', s_2)  \cdot e_1(P, s_1) + 
       {F_2(q^2) \over 2M^2} \left[ e_2^*(P', s_2) \cdot q \right] \left[ 
       e_1(P, s_1) \cdot q \right] \right\} \nonumber \\
       & + & F_3(q^2) \left\{ \left[ e_2^{\mu}(P', s_2) \right]^* \left[ 
       e_1(P, s_1) \cdot q \right] - e_1^{\mu}(P, s_1) \left[ e_2^*(P', s_2) 
       \cdot q \right]  \right\} \nonumber \\
       & + & e_{2 \alpha}^*(P', s_2) ~ B^{\mu, \alpha \beta}(P,  P', \omega) 
       ~ e_{1 \beta}(P, s_1) ~ , 
       \label{eq:F}
 \ee
which means that only the form factors $F_i(q^2)$ are the physical ones describing the elastic response for spin-1 systems \footnote{In principle the form factors $H_j(q^2)$ can be eliminated by subtracting from the triangle diagram the contributions of self-energy diagrams. Note that in Eq. (\ref{eq:F+H}) the term containing the form factor $H_4(q^2)$ takes into account possible loss of gauge invariance in the triangle diagram, which can be still compatible with time reversal symmetry for spin-1 systems, at variance with the case of spin-0 systems.}. 

\indent Let us consider the Breit frame where $q = (q^-, q^+, q_x, q_y) = (0, 0, Q, 0)$ with $Q^2 \equiv q_{\perp}^2 = -q^2 \geq 0$, $P = (P^-, P^+, -Q/2, 0)$, $P' = ({P'}^-, {P'}^+, Q/2, 0)$ and ${P'}^- = {P'}^+ = P^- = P^+ = \sqrt{M^2 + Q^2 / 4}$. In such a frame the independent {\em good} components with $\mu = +$ are four, namely $(\alpha, \beta) = (y, y), (x, x), (+, +), (x, +)$, and the {\em good} ones with $\mu = y$ are two, namely $(\alpha, \beta) = (x, y), (+, y)$ \footnote{As a matter of fact, in our Breit frame one has $I^{+, +x} = - I^{+, x+}$, $I^{y, yx} = - I^{y, xy}$ and $I^{y, +y} = I^{y, y+}$, while all the other {\em good} components are identically vanishing.}. Explicitly, one has
 \be
         I^{+, yy} & = & 2P^+ F_1(Q^2) ~, \nonumber \\
         I^{+, xx} & = & 2P^+ \left[ (1 + 2 \eta) (1 + \eta)  F_1(Q^2) - 2 
         \eta (1 + \eta)^2 F_2(Q^2) - 2\eta F_3(Q^2) \right. 
         \nonumber \\
         & + & \left. \eta H_1(Q^2) + 2\eta H_2(Q^2) \right] ~, \nonumber \\
         I^{+, ++} & = & 2P^+ (1 + \eta) \left[ (1 - 2\eta) F_1(Q^2) + 
         2\eta^2 F_2(Q^2) + 2\eta F_3(Q^2) - H_1(Q^2) \right] ~, 
         \nonumber \\
         I^{+, x+} & = & 2P^+ \sqrt{\eta(1 + \eta)} \left[ -(1 + 2\eta) 
         F_1(Q^2) + 2\eta (1 + \eta) F_2(Q^2) + (1 + 2\eta) F_3(Q^2) \right. 
         \nonumber \\
         & - & \left. H_1(Q^2) - H_2(Q^2) \right] ~, \nonumber \\
         I^{y, xy} & = & Q \left[ (1 + \eta F_3(Q^2) - {1 \over 2} H_3(Q^2) 
         \right] ~, \nonumber \\
         I^{y, +y} & = & Q \left[ \sqrt{\eta (1 + \eta)} F_3(Q^2) - {1 \over 
        2} \sqrt{ {1 + \eta \over \eta}} H_3(Q^2) \right]
        \label{eq:good}
 \ee
where $\eta \equiv Q^2 / 4M^2$. Thus, we have six equations for the six form factors $F_i(Q^2)$ and $H_i(Q^2)$ with $i = 1, 2, 3$ \footnote{The form factor $H_4(q^2)$, which is related to possible loss of gauge invariance in the triangle diagram, can be determined only from the $\mu = x$ components of the tensor (\ref{eq:F+H}); more precisely, in our Breit frame the only non-vanishing component with $\mu = x$ is $I^{x, x+} =  I^{x, +x}$. Explicitly one has $I^{x, x+} = P^+ \{H_3(q^2) + 4 \eta H_4(q^2) \}$.}. From the last two equations in (\ref{eq:good}), which derive from $\mu = y$, we can get the form factors $F_3(Q^2)$ and $H_3(Q^2)$; then, after substitution in the first four equations,  which derive from $\mu = +$, the other four form factors can be obtained. Explicitly for the physical form factors one has
 \be
       F_1(Q^2) & = & {I^{+, yy} \over 2P^+} ~, \nonumber \\
       F_2(Q^2) & = & {1 \over 2\eta} {I^{+, yy} - I^{+, xx} \over 2P^+} + 
       {1 \over 2 (1 + \eta)} {I^{+, ++} \over 2P^+} - {1 \over \sqrt{\eta 
       (1 + \eta)}} {I^{+, x+} \over 2P^+} ~, 
       \nonumber \\
       F_3(Q^2) & = & {I^{y, xy} \over Q} - \sqrt{{1 + \eta \over \eta}} 
       {I^{y, +y} \over Q} ~.
       \label{eq:solution}
 \ee

\indent At variance with the spin-0 case the {\em good} components of the tensor (\ref{eq:spin1}) with $\mu = y$ are essential for the extraction of the physical form factors, more precisely for the determination of $F_3(Q^2)$. Note that the form factors $F_1(Q^2)$ and $F_2(Q^2)$ are determined only by the {\em good} components with $\mu = +$.

\indent We want to stress that by means of Eq. (\ref{eq:solution}) the angular condition problem for spin-1 systems is completely overcome and the extraction of the physical form factors is not plagued at all by spurious effects related to the loss of rotational covariance. Therefore, a physically meaningful one-body approximation, in which all the constituents are on their mass-shells, can be consistently formulated in the limit $q^+ = 0$.

\indent We are now in the right position to overcome the negative result of Ref. \cite{Sauer} that for spin-1 systems the matrix elements of the $(+)$ component of the current are affected by the longitudinal zero mode. As already pointed out, the presence of a residual effect from the $Z$-graph in the limit $q^+ = 0$ is totally due to the specific choice (\ref{eq:choice_Fre}) for the regularizing functions $\Lambda_i$. For such a choice the longitudinal zero-mode can affect the components of the tensor (\ref{eq:triangle_1}) having at least one of the indexes $\alpha$, $\beta$ and $\mu$ equal to $(-)$. Therefore, since the relation between the matrix elements of the current and the tensor (\ref{eq:triangle_1}) is given by 
 \be
       J_{s_2, s_1}^{\mu}(P_1, P_2) = e_{2 \alpha}^*(P_2, s_2) ~ 
       {\cal{J}}_F^{\mu, \alpha \beta}(P_1,  P_2) ~ e_{1 \beta}(P_1, s_1) ~,
       \label{eq:current}
 \ee
it is possible that in the contraction appearing in the r.h.s. of Eq. (\ref{eq:current}) the {\em bad} components with $\alpha$ and/or $\beta$ equal to $(-)$ are involved even if we choose $\mu = +$. When this happens (as it is the case of the longitudinal polarization state $s_{1(2)} = 0$), the $(+)$ component of the current (\ref{eq:current}) is affected by the $Z$-graph even in the limit $q^+ = 0$. The same results hold as well for $\mu \neq +$, so that all the current components (\ref{eq:current}) may be affected by the $Z$-graph in the limit $q^+ = 0$. Therefore, our recipe of employing only the {\em good} components of the tensor (\ref{eq:triangle_1}) works well also in case of the regularizing functions (\ref{eq:choice_Fre}) adopted in Ref. \cite{Sauer} and allows to avoid any residual effect of the $Z$-graph on the extraction of the form factors of spin-1 systems in the limit $q^+ = 0$.

\section{Application to the $\rho$-meson}

\indent In this Section we apply our basic Eq. (\ref{eq:solution}) to the calculation of the elastic form factors of the $\rho$-meson. Our aim is not to provide a full $QCD$-motivated calculations of such form factors, which would require to go beyond the simple constituent quark picture. Instead of that we want simply to compare our covariant results with the non-covariant ones obtained in Ref. \cite{CAR_rho} using a $ LF$ approach where the angular condition is not satisfied and the extraction of the form factors is therefore plagued by spurious effects related to the loss of rotational covariance (see also Ref. \cite{Volodia}).

\indent The structure of the $LF$ wave function for an $S$-wave vector system reads as
 \be
        | V_1; \beta \rangle_{LF} = \mbox{R}^{(1)}(\xi, \vec{k}_{\perp}; 
        \beta) w(k) \sqrt{{A(\xi, \vec{k}_{\perp}) \over 4\pi}}
        \label{eq:LFwf_1}
 \ee
where $w(k)$ is the $S-wave$ radial wave function. In terms of Dirac spinors the spin matrix $\mbox{R}^{(1)}(\xi, \vec{k}_{\perp}; \beta)$ can be written as
 \be
       \left[ \mbox{R}^{(1)}(\xi, \vec{k}_{\perp}; \beta) \right]_{\lambda_1 
       \lambda} = {1 \over \sqrt{2}} {1 \over 
       \sqrt{M_{10}^2 - (m_I - m)^2}} \bar{u}(\tilde{p}_1, \lambda_1) 
       \tilde{V}^{\beta} v(\tilde{p}, \lambda) ~.
       \label{eq:Melosh_1}
 \ee
with
 \be
       \tilde{V}^{\beta} = \gamma^{\beta} - {(\tilde{p}_1 - 
       \tilde{p})^{\beta} \over M_{10} + m_I + m} ~.
       \label{eq:V_tilde}
 \ee

\indent Adopting the $LF$ wave function (\ref{eq:LFwf_1}) and the one-body approximation (\ref{eq:one-body}) for the $e.m.$ current, we define the $LF$ tensor $I_{LF}^{\mu, \alpha \beta}$ as
 \be
       I_{LF}^{\mu, \alpha \beta} \equiv ~ 2P_1^+ ~ _{LF}\langle V_2; \alpha 
       | \bar{u}_2 J_1^{\mu} u_1 | V_1; \beta \rangle_{LF}
       \label{eq:tensor_LF} ~ .
 \ee
For $\beta \neq -$ one has $\bar{u}(\tilde{p}_1, \lambda_1) \tilde{V}^{\beta} v(\tilde{p}, \lambda) = \bar{u}(\tilde{p}_1, \lambda_1) \overline{V}^{\beta} v(\tilde{p}, \lambda)$, and therefore the components of $I_{LF}^{\mu, \alpha \beta}$ having $\mu = +, \perp$ and $\alpha, \beta \neq -$ coincide with the corresponding components of the tensor $I^{\mu, \alpha \beta}$ [i.e., the {\em good} components of the tensor (\ref{eq:sp_1})] obtained using the modified bound-state vertexes (\ref{eq:V_new}) with the regularizing functions $\Lambda_i$ given by Eq. (\ref{eq:radial}) with $w_1(k) = w_2(k) = w(k)$. Thus, adopting the Breit frame described in the previous Section, the physical form factors $F_1(Q^2), F_2(Q^2)$ and $F_3(Q^2)$ can be uniquely extracted from the $LF$ tensor (\ref{eq:tensor_LF}) by means of Eq. (\ref{eq:solution}).

\indent In the $LF$ approach adopted in Ref. \cite{CAR_rho}, only the matrix elements of the $(+)$ component of the $e.m.$ current are considered, viz.
 \be
       J_{s_2, s_1}^+= e_{2 \alpha}^*(P_2, s_2) ~ {\cal{J}}_{LF}^{+, \alpha 
       \beta}(P_1,  P_2) ~ e_{1 \beta}(P_1, s_1) ~,
       \label{eq:current_LF}
 \ee
After considering general symmetry properties of the current operator (like, e.g., time reversal symmetry) the number of independent matrix elements (\ref{eq:current_LF}) turns out to be four, while the physical form factors are three. A further condition arises from the rotational invariance of the charge density, which however involves transformations based upon Poincar\'e generators depending on the interaction. Such an additional constraint is known as the angular condition which takes for a vector system the following form \cite{GK}
 \be
       \Delta(Q^2) \equiv (1 + 2 \eta) J_{1 1}^+ + J_{1 -1}^+ - \sqrt{8 
       \eta} J_{1 0}^+ - J_{0 0}^+ = 0
       \label{eq:angular}
 \ee
where $\eta \equiv Q^2 / 4 M^2$. According to our analysis the matrix elements $J_{1 0}^+$ and $J_{0 0}^+$ are not {\em good} amplitudes since they receive the admixture of the {\em bad} components of the tensor (\ref{eq:sp_1}) after contraction with the polarization four-vectors. Therefore, the angular condition (\ref{eq:angular}) is not satisfied by the matrix elements of the one-body $e.m.$ current (\ref{eq:one-body}). Various prescriptions have been proposed in the literature, but the final result depends upon the specific prescription and, consequently, the determination of the physical form factors $F_i(Q^2)$ is not unique.

\begin{figure}[htb]

\vspace{0.25cm}

\centerline{\epsfxsize=16cm \epsfig{file=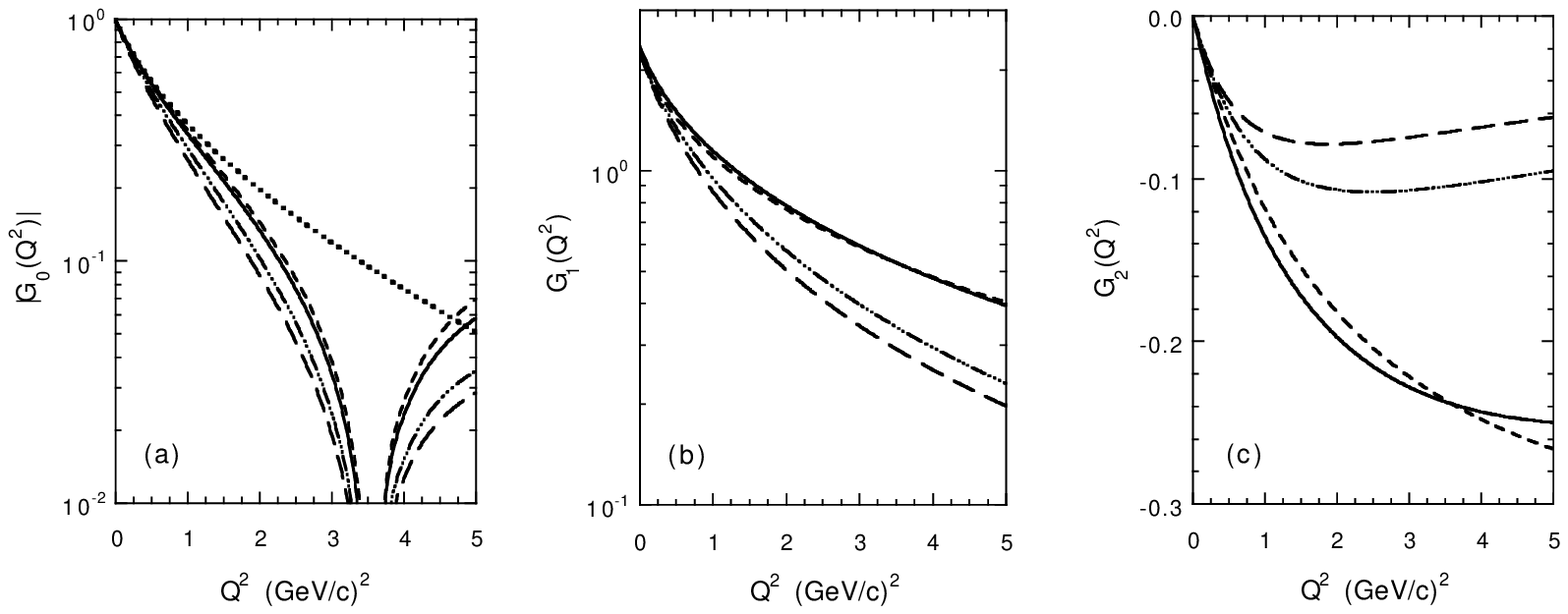}}

\vspace{0.25cm}

{\small \noindent Figure 3. Elastic form factors of the $\rho$-meson, $|G_0(Q^2)|$ (a),  $G_1(Q^2)$ (b) and $G_2(Q^2)$ (c) [see Eqs. (\ref{eq:Gi})], versus the squared four-momentum transfer $Q^2$, assuming point-like constituent quarks [i.e, $f_1^{(j)} = e_j$ and $f_2^{(j)} = 0$ in Eq. (\ref{eq:one-body})]. Solid lines: covariant $LF$ approach of the present work, given by Eq. (\ref{eq:solution}). The dotted, short-dashed, triple-dot-dashed and long-dashed lines are the results obtained in Ref. \cite{CAR_rho} within the $FFS$, $GK$, $CCKP$ and $BH$ prescriptions, respectively. In (b) and (c) the results of the $FFS$ prescription coincide with the $CCKP$ ones. As for the radial wave function $w(k)$, appearing in Eq. (\ref{eq:LFwf_1}), the $\rho$-meson eigenfunction of the quark potential model of Ref. \cite{GI} has been adopted.}

\vspace{0.25cm}

\end{figure}

\indent In case of the $\rho$-meson \cite{CAR_rho}  four prescriptions were explicitly considered and labeled as $GK$ from Ref. \cite{GK}, $FFS$ from Ref. \cite{FFS}, $CCKP$ form Ref. \cite{CCKP} and $BH$ form Ref. \cite{BH}. The expressions of the form factors within the four prescriptions can be read off from Eqs. (5-8) of Ref. \cite{CAR_rho}. In terms of the conventional charge $G_0(Q^2)$, magnetic $G_1(Q^2)$ and quadrupole $G_2(Q^2)$ form factors, defined as
 \be
       G_0(Q^2) & = & F_1(Q^2) + {2 \eta \over 3} \left[ F_1(Q^2) - F_3(Q^2) 
       - (1 + \eta) F_2(Q^2) \right] \nonumber \\
       G_1(Q^2) & = & F_3(Q^2) \nonumber \\
       G_2(Q^2) & = & {\sqrt{8} \eta \over 3} \left[ F_1(Q^2) - F_3(Q^2) 
       - (1 + \eta) F_2(Q^2) \right] ~,
       \label{eq:Gi}
 \ee
the results of Ref. \cite{CAR_rho} are reported in Fig. 3 and compared with our covariant calculation (solid lines) obtained through Eq. (\ref{eq:solution}). It can clearly be seen that all form factors calculated within the $FFS$, $CCKP$ and $BH$ prescriptions strongly deviates from our covariant result, while the effects of the violation of the angular condition seems to be minimized in the $GK$ prescription. Finally, as for the magnetic, $\mu_V = G_1(Q^2 = 0)$, and the quadrupole, $Q_V = lim_{Q^2 \to 0} (3 \sqrt{2} / Q^2) G_2(Q^2)$, moments the results of Ref. \cite{CAR_rho} were the same in all the four prescriptions considered,  and equal to $\mu_V = 2.26$ and $Q_V = -0.024 ~ fm^2$ in case of the quark potential model of Ref. \cite{GI}. Although such results do not depend on the specific prescription adopted, they are still affected by spurious effects related to the violation of the rotational covariance. As a matter of fact, the results obtained within our covariant $LF$ approach, which we stress are not affected at all by the angular condition, are: $\mu_V = 2.35$ and $Q_V = -0.031 ~ fm^2$. For the squared charge radius of the $\rho$-meson we get $r_{ch}^2 \equiv - 6 [dG_0(Q^2) / dQ^2 ]_{Q^2 = 0} = 0.33 ~ fm^2$ assuming point-like constituent quarks.

\section{Conclusions}

\indent In this work we have addressed the issue of the connection between the Feynman triangle diagram and the Hamiltonian light-front formalism for spin-0 and spin-1 two-fermion systems. The most important result we have achieved is that in the limit $q^+ = 0$ the physical form factors for both spin-0 and spin-1 systems can be uniquely determined using only the {\em good} amplitudes which are not affected by spurious effects related to the loss of rotational covariance present in the light-front formalism. At the same time, the unique feature of the suppression of the pair creation process is maintained, so that a physically meaningful one-body approximation, in which all the constituents are on their mass-shells, has been consistently formulated in the limit $q^+ = 0$.

\indent We have given a new definition of {\em good} amplitudes to be used for extracting the physical form factors. For the reference frame specified by the relations $q^+ = 0$ and $q_x = q_{\perp}$, $q_y = 0$ in the transverse $(x,y)$-plane, the good amplitudes correspond to $\mu = +$ and $\mu = y$ for both spin-0 and spin-1 systems. For spin-1 system one should in addition use $\alpha, \beta \neq (-)$ for the components of the tensor ${\cal{J}}_F^{\mu, \alpha \beta}$ [see Eq. (\ref{eq:triangle_1})]. The good amplitudes defined in this way contain only physical form factors.

\indent According to this definition, for spin-1 systems the $(+)$ component of the one-body current used in Refs. \cite{Frederico_rho,Sauer} is not a good amplitude, because it contains the admixture of bad amplitudes after contraction with the longitudinal polarization vectors. Therefore, our procedure to extract the physical form factors guarantees that the latter are not affected by any residual effect due to the pair creation process in the limit $q^+ = 0$. 

\indent We have also shown that by means of a specific choice of the off-shell behavior of the bound-state vertex the Feynman triangle diagram and the one-body form factors of the standard light-front formalism match exactly. To this end we have explicitly constructed a model for the $\omega$-dependent bound-state wave function, which takes into account the suppression of the endpoint regions (in accordance with the general properties of the wave functions, valid also in $QCD$), and which guarantees at the same time the suppression of any off-shell effect of the active constituents (in particular, of their instantaneous propagation).

\indent The {\em on-shell} part of the bound-state vertex corresponds to the usual light-front wave function and this makes it possible to establish a very useful link with potential models. We stress that such a feature is very important for phenomenological applications, particularly in case of quark models of the hadron structure.

\indent We have applied our basic Eq. (\ref{eq:solution}) to the case of the $\rho$ meson for comparison with non-covariant light-front results available in the literature \cite{CAR_rho}. The calculation of the deuteron elastic form factors is in progress and the results will be published elsewhere.

\indent Before concluding we remind that the limit $q^+ = 0$ is possible only for space-like $q$ ($q^2 \leq 0$). Indeed, for time-like $q$ ($q^2 > 0$) one has always $q^+ \neq 0$. In this case one needs to perform an analytic continuation of the Feynman triangle diagram from space-like to time-like $q$. This cannot be easily done in the light-front formalism because the contribution of the $Z$-graph cannot be eliminated when $q^2 > 0$. However, the proper analytic continuation of the Feynman triangle diagram can be achieved by means of the so-called dispersion approach, which is described in Ref. \cite{Dima} and has been extensively applied to time-like processes, like heavy meson weak decays, in Ref. \cite{MNS}. We stress that for space-like $q$ the dispersion approach result matches the light-front one (see Ref. \cite{Dima}).

\section*{Acknowledgments} The authors gratefully acknowledge T. Frederico, V.A. Karmanov and W. Polyzou for many valuable discussions and for a critical reading of the manuscript.  One of the authors, D. M., would like to thank the Alexander von Humboldt-Stiftung for financial support.

\end{document}